\documentclass[amsmath,amssymb,showpacs,twocolumn,pra]{revtex4-1} 
\usepackage{color}
\usepackage{graphicx}
\usepackage{amssymb}
\usepackage{amsmath}
\usepackage{mathrsfs}
\usepackage{subfigure}
\usepackage{hyperref}
\usepackage{breakurl}

\newcommand{\ud}{\mathrm{d}}

\begin{document}
\title{Oscillatory dynamics in nano-cavities with non-instantaneous Kerr response}
\author{ Andrea Armaroli}
\email{andrea.armaroli@unife.it}
\affiliation{Department of Engineering, University of Ferrara, Via Saragat 1, 44122 Ferrara, Italy}
\author{Stefania Malaguti}
\affiliation{Department of Engineering, University of Ferrara, Via Saragat 1, 44122 Ferrara, Italy}
\author{Gaetano Bellanca}
\affiliation{Department of Engineering, University of Ferrara, Via Saragat 1, 44122 Ferrara, Italy}
\author{Alfredo de Rossi and Sylvain Combri\'{e}}
\affiliation{Thales Research and Technology, Palaiseau cedex, 91767 France}
\author{Stefano Trillo}
\affiliation{Department of Engineering, University of Ferrara, Via Saragat 1, 44122 Ferrara, Italy}

\date{\today}
\begin{abstract}
We investigate the impact of a finite response time of Kerr nonlinearities over the onset of spontaneous oscillations (self-pulsing)
occurring in a nanocavity. The complete characterization of the underlying Hopf bifurcation in the full parameter space allows us
to show the existence of a critical value of the response time and to envisage different regimes of competition with bistability.
The transition from a stable oscillatory state to chaos is found to occur only in cavities which are detuned far off-resonance, 
which turns out to be mutually exclusive with the region where the cavity can operate as a bistable switch.
\end{abstract}
\pacs{42.65.Pc,42.65.Sf,42.55.Sa}
\maketitle
\section{INTRODUCTION}
Self-pulsing (SP), the onset of spontaneous oscillations, is a universal feature of nonlinear structures with feedback.
As long as passive systems are concerned SP has been investigated theoretically in settings ranging from isolated
ring cavities \cite{Ikeda80,Ikeda82,Lugiato82} and parametric intracavity mixing \cite{SW83,L88,TH96} to Bragg gratings \cite{WC82,P07} or grating-assisted backward frequency conversion schemes \cite{D95,C05}, and it is still a subject of active research \cite{Maes09,GB11,Malaguti11}.
In particular, the dynamics of nonlinear passive cavities, whose study has been pioneered in the eighties \cite{Ikeda80,Ikeda82,Lugiato82}, is extremely rich
encompassing stable as well as chaotic SP which can compete with bistabilities and transverse effects \cite{LL87}. 
Historically, SP and chaos (the optical equivalent of strong turbulence) 
have been first analyzed by means of delay-differential models  accounting for the round-trip delay at each passage, which can be large in ring cavities. 
An instability named after Ikeda occurs in this framework when the relaxation time of the nonlinear response is much shorter than the transit time \cite{Ikeda80} and 
has been tested experimentally \cite{expSP}.
However SP and so-called weak turbulence occurs also in the opposite (say {\em short cavity}) limit,
where the delay can be averaged out to end up with a differential mean-field model \cite{Ikeda82}. 
This regime becomes important nowadays where nano-cavities are employed for many modern photonic applications \cite{Vahala03},
including bistability \cite{MIT02}, demonstrated in photonic crystal (PhC) membranes 
which offer great flexibility of design as well as high nonlinear performances in semiconductors
\cite{Barclay05,Notomi05,Uesugi06,Weidner06,Weidner07,Combrie08,deRossi09,Yang07}.
In these cavities, transverse effects are absent and their dimensions are so small that the response time of the medium can be much larger than the light transit time in the cavity, 
yet being comparable with the photon lifetime which is strongly enhanced on account of the large quality factor $Q$.
SP in such nano-cavities has been recently predicted, owing to the free-carrier dispersion induced by two-photon absorption \cite{Malaguti11}.
The main features of such mechanism is the existence of a critical value of the time relaxation constant $\tau$, 
as well as a wide region of the parameter space where stable (non-chaotic) SP can be potentially observed. 
In this paper we analyse the dynamics of a nano-resonator in the good-cavity limit when the underlying nonlinear mechanism is a Kerr-like nonlinearity with finite response time.
Our analysis is based on the differential model proposed in Ref.~\cite{Ikeda82}. In spite of the simplicity of such model, which
makes it an ideal prototype for understanding the role of relaxation processes, a full characterization of SP and its competition with bistability in the full parameter space 
was never reported (to the best of our knowledge) after Ref.~\cite{Ikeda82}.
We propose such analysis, adopting a different normalization with respect to that employed in Ref. \cite{Ikeda82}, better aimed at capturing
the key role of the relaxation time. This is especially important nowadays in view of assessing how a given designed nano-cavity may be expected to behave 
by changing the characteristic relaxation time of the nonlinearity as a consequence of choosing different materials and/or adopting
techniques for fine tuning their response time. We propose an analytical characterization of the SP instability and its competition
with bistability in the full parameter space, pointing out the qualitative similarity with the features observed for a nonlinearity dominated by free-carrier dispersion \cite{Malaguti11}.
Nonetheless, we further investigate  also the destabilization mechanism of the oscillatory states initially described in Ref.~\cite{Ikeda82}, 
showing that the chaotic regime, being confined to far off-resonance cavities, is indeed mutually exclusive with bistable switching.

\section{Model definition and linear stability analysis}

We start from the following dimensionless coupled-mode model that rules the temporal evolution of the normalized
intra-cavity field $a(t)$ coupled to the frequency deviation $n(t)$, owing to the intensity-dependent refractive index change
\begin{subequations} \label{eq:model}
\begin{align}
&\frac{\ud a}{\ud t} = \sqrt{P} + i (\delta + \chi n) a - a,\label{eq1}\\
\tau &\frac{\ud n}{\ud t} + n = |a|^2.\label{eq2}
\end{align}
\end{subequations}
Noteworthy Eqs. \eqref{eq:model} describe a photonic crystal nano-cavity with high Q coupled to a line-defect waveguide \cite{Barclay05,Uesugi06,deRossi09}.
They implicitly assume that the nonlinearity is dominated solely by the Kerr effect with relaxation time $\tau$, 
while other possible nonlinear contributions, e.g. two-photon absorption along with the free-carrier dispersion \cite{Barclay05,Uesugi06,deRossi09,Malaguti11}), are neglected.
Here $P$ is the normalized power injected in the cavity through coupling with the waveguide, and $|a|^2$ is the normalized intra-cavity energy,
which can be easily rescaled into real-world units by comparison with widely used dimensional models (see, e.g., Ref.~\cite{deRossi09}). 
It is worth emphasizing that a unit coefficient in front of the loss term in Eqs.~\eqref{eq:model} implies that the time 
$t$ is measured in units of the inverse damping coefficient $1/\Gamma_0=2Q/\omega_0=2 t_c$, where $Q$, $\omega_0$, and $t_c$
stand for the overall quality factor, the resonant frequency, and the photon lifetime of the cavity, respectively.
In these units, the two key (normalized) parameters are the detuning  $\delta  = (\omega_0 - \omega) / \Gamma_0$, 
and the time constant $\tau  = \tau_p \Gamma_0$,  where $\tau_p$ is the response time of the nonlinearity in real-world units,
while $\chi=\pm 1$ accounts for the sign of the nonlinear Kerr coefficient. 
We point out that our model differs from \cite{Ikeda82}, inasmuch as the time scale is referred to the cavity lifetime instead of the response time of the medium.
Indeed Eqs.~\eqref{eq:model} can be reduced to the model analyzed in Ref. \cite{Ikeda82} by means of the substitution
$a, n, t \rightarrow a/\sqrt{\tau}, n/\tau, \tau t$. The effect of such transformation is however to rescale the detuning and the injected power
in such a way that they become dependent on the response time of the medium, which is not suitable for our purpose
of investigating the impact of the relaxation time on the dynamics of a given cavity with fixed characteristics.

For a cw driving $P=constant$, Eqs.~\eqref{eq:model} have the following steady-state solution $a(t)=A$, $n(t)=N=|A|^2$, where
\begin{equation}\label{steady}
P = E \left[ (1 +(\delta +\chi E)^2 \right],
\end{equation}
$E=|A|^2$ being the stationary intra-cavity energy. 
It is well known that bistability occurs for $\delta > \sqrt{3}$ when $\chi=-1$, and $\delta < -\sqrt{3}$ when $\chi=1$ \cite{LL87}. 
In the discussion below,  we will focus on the latter case, where the cavity resonance is blue-shifted due to the 
nonlinearity, a case which is directly comparable with the net effect of free-carrier dispersion induced by two-photon absorption  \cite{Malaguti11}.
All the conclusions of this paper remain valid also for $\chi=-1$, provided $\delta \rightarrow -\delta$.
The values of intracavity energies corresponding to the knees of the bistable response are
\begin{equation}\label{knees}
E_b^{\pm} = \frac{-2\chi \delta \pm \sqrt{ \delta^2-3}}{3},
\end{equation}
and the corresponding input powers $P_b^{\pm}=P(E_b^{\pm})$ can be calculated by means of Eq.~(\ref{steady}).

The stability of the solution (\ref{steady}) can be investigated by plugging into
Eqs.~\eqref{eq:model} the ansatz $a(t)=A+\delta a(t)$, $n(t)=N+\delta n(t)$, 
while retaining linear terms in the perturbations $\delta a, \delta n$. 

The perturbation column array  $\varepsilon \equiv (\delta a, \delta a^*, \delta n)^T$  is found to obey the following linearized equation
\begin{subequations}
\label{eq:lin}
\begin{align}
	\frac{\ud \varepsilon}{\ud t} &=M \varepsilon; \label{lsa1}\\
	M &= 
	\begin{pmatrix}
			i\hat{\delta} -1 & 0  &   i\chi A\\
		 0 & -i\hat{\delta} -1  &   -i\chi A^* \\
 		 A^*/\tau &  A/\tau  &   -1/\tau
	\end{pmatrix}, \label{lsa2}
\end{align}
\end{subequations}
where $\hat{\delta} \equiv \delta + \chi E$.  

The characteristic equation of $M$  reads as
\begin{equation}
\lambda^3 + a_2 \lambda^2 + a_1 \lambda + a_0=0,
\end{equation}
where the coefficients are 
$a_2=2 +\frac{1}{\tau}$, $a_1=\left( 1 + \hat{\delta}^2 + \frac{2}{\tau} \right)$, and 
$a_0 = \frac{1}{\tau} \left( 1 + \hat{\delta}^2 + 2 \chi E \hat{\delta} \right)$. 
\begin{figure}[ht]
\centering
\includegraphics[width=9cm]{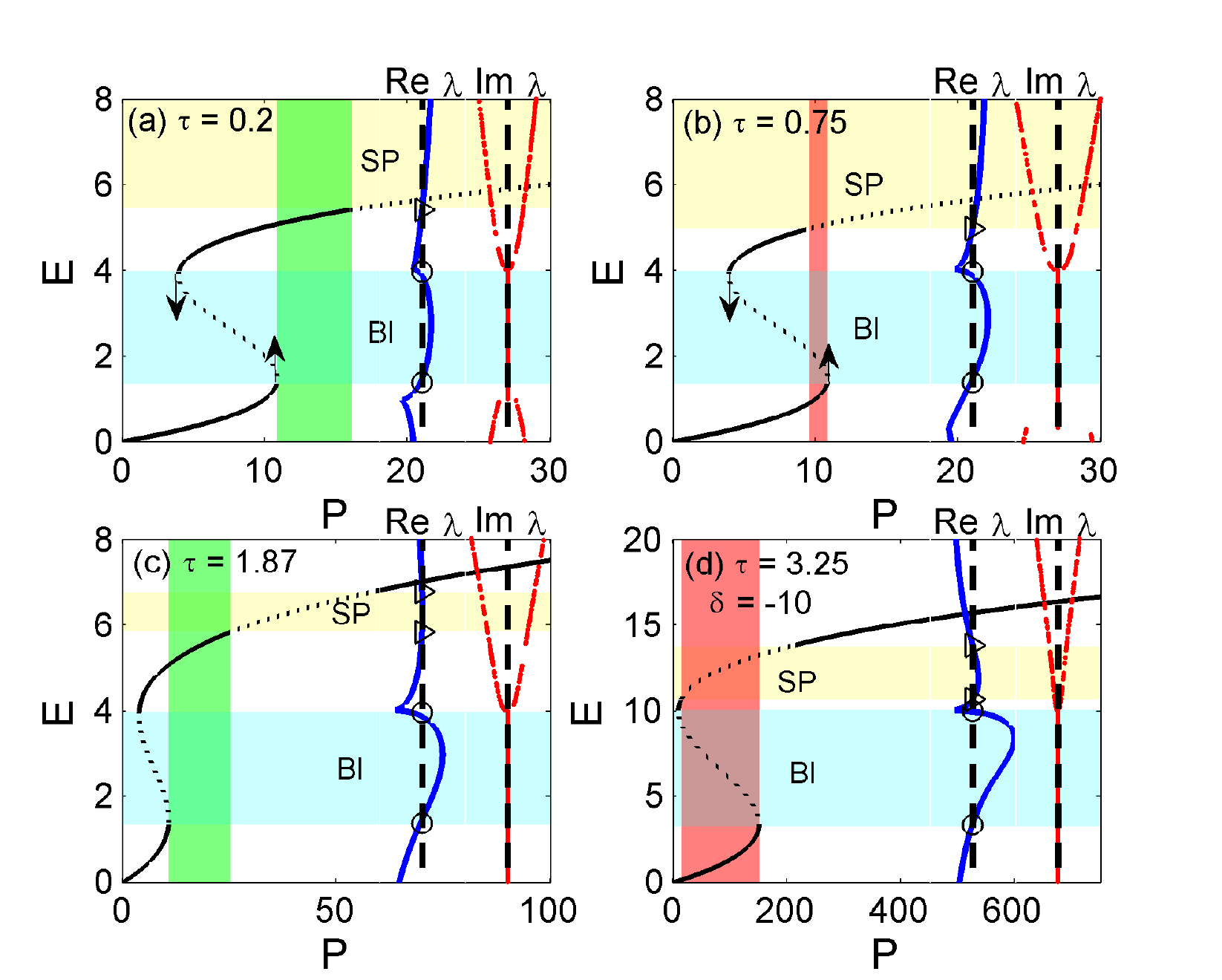}
\caption{(Color online) Steady-state response $E$ vs.~$P$ for $\delta=-4$ (a-b-c) and  $\delta=-10$ (d) and different $\tau$ ($\chi=1$).
Stable and unstable branches are reported as solid and dotted lines, respectively. The blue and red curves superimposed on the right side
show how the real part $Re(\lambda)$ and imaginary part $Im(\lambda)$ (of the dominant eigenvalue underlying the instability) change with $E$.
Bifurcation points are highlighted over with the dashed line $Re(\lambda)=0$.
The shaded regions labeled BI (light blue) and SP (light yellow) correspond to the negative slope branch of the bistable response (real eigenvalue)
and SP instability (pair of conjugate eigenvalues with positive real part), respectively. Four different scenarios are shown:
(a) SP occurs at $P=P_H^-$ values above the bistable knee value $P_b^+$, and is unbounded for increasing $P$; 
(b) SP occurs at $P=P_H^-$ values below the bistable knee $P_b^+$, still being unbounded;
(c) as in (a), except SP occurs in a finite range below a given value $P=P_H^+$; 
(d) as in (b), except SP occurs in a finite range below a given value $P=P_H^+$.
The shaded green regions yield the range of power where bistable up-switching to a stable state is permitted, 
while red ones identify the coexistence of a SP and an unstable saddle branch.
} 
\label{fig:eigs}\end{figure}

SP occurs when the system undergoes a Hopf bifurcation, i.e.~a pair of complex conjugate eigenvalues $\lambda_R \pm i \lambda_I$ 
crosses into the right half complex plane, entailing an exponential growth of a pulsating perturbation with period $T=2\pi/ |\lambda_I |$.
The bifurcation point ($\lambda_R=0$) corresponds to the constraint $a_1 a_2 = a_0$, which can be solved to yield
 the following explicit expression for the SP (Hopf) threshold values $E_H^{\pm}$
\begin{equation} \label{eq:Hopf}
E_H^{\pm} = \frac{-\chi \delta  \left( 2 - \frac{1}{\tau} \right) \pm 
\sqrt{\frac{\delta^2}{\tau^2}-4 \left(1 + \frac{1}{\tau} \right)^2 \left(1 - \frac{1}{\tau}\right)}}{2\left(1 - \frac{1}{\tau} \right)},
\end{equation}
and the corresponding injected power threshold $P_H^{\pm} = E_H^{\pm}  \left[ (1 +(\delta +\chi E_H^{\pm} )^2 \right]$.

\begin{figure}[h!] 
\centering
\includegraphics[width=8cm]{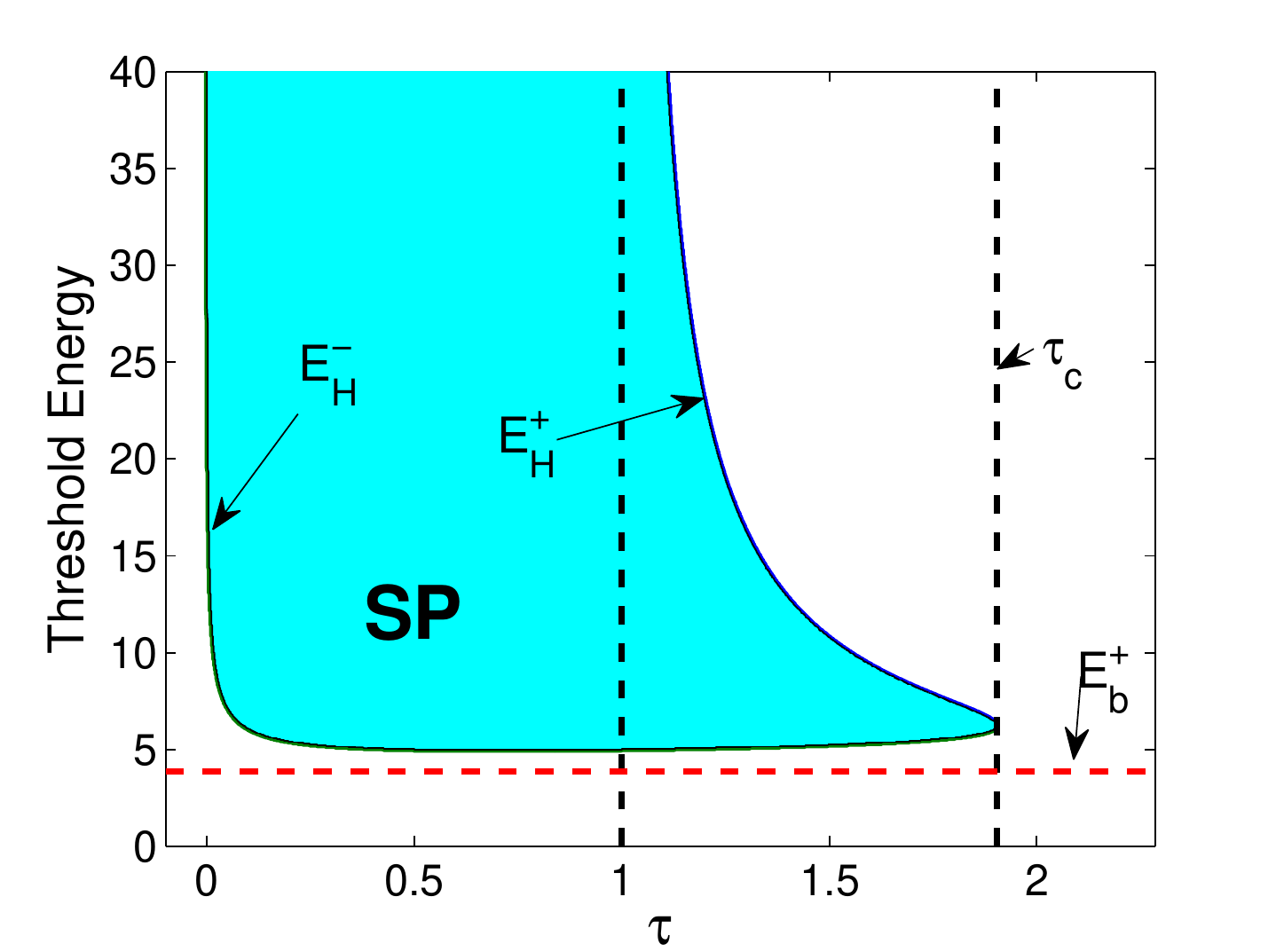}
\caption{(Color online) SP threshold energies $E_H^{\pm}$ as a function of $\tau$ for fixed detuning $\delta = -4$.
The shaded area corresponds to the domain $E_H^{-} \le E \le E_H^{+}$ where SP occurs,
which lies above the upper knee level of energy $E_b^+$ (red dashed line).
} 
\label{fig:th}\end{figure}
\begin{figure} 
\centering
\includegraphics[width=0.45\textwidth]{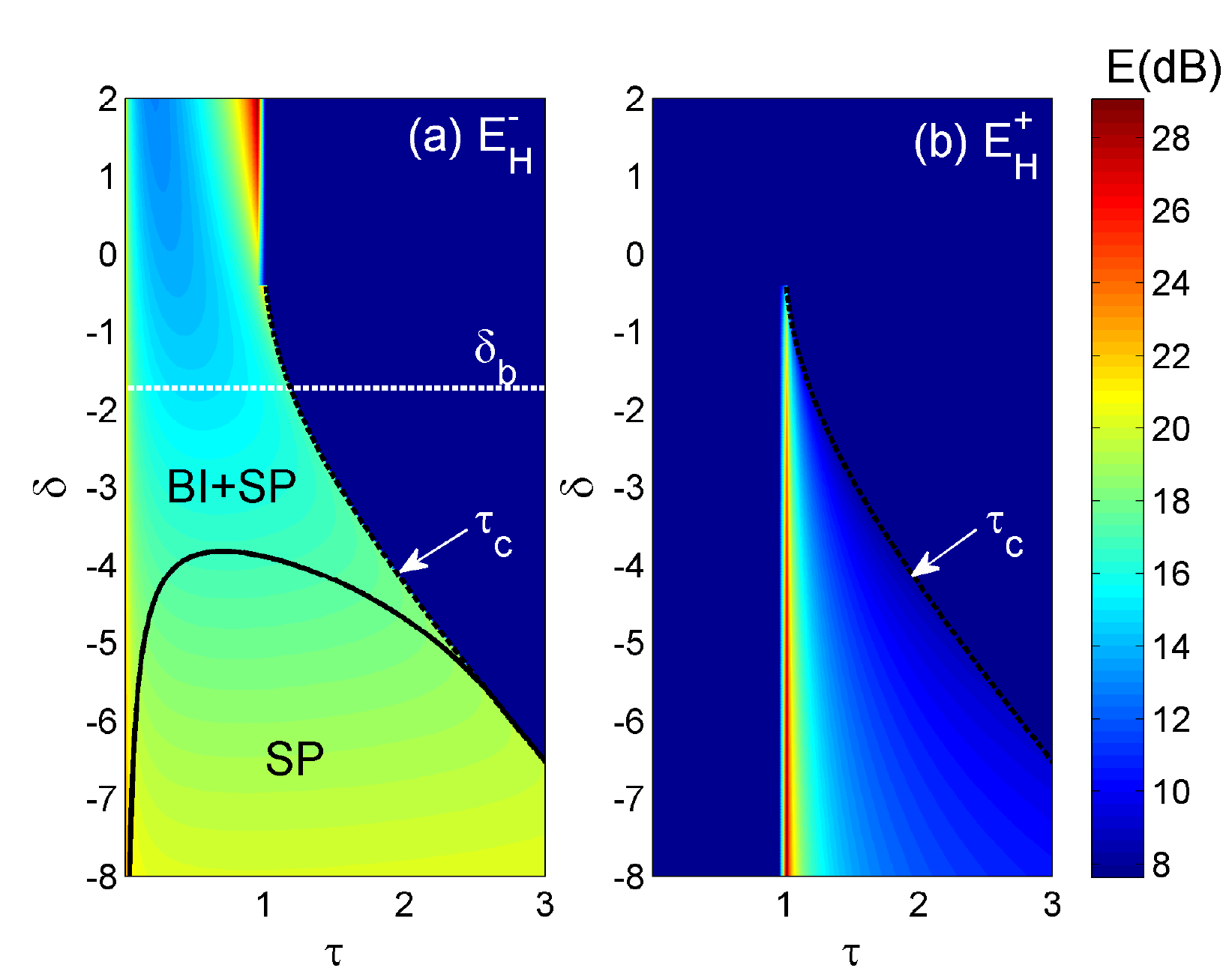}
\caption{(Color online) Color level plots of the (a) "on" $E_H^{-}$ and (b) "off" $E_H^{+}$ values
of threshold energy for SP, in the parameter plane ($\tau, \delta$). Bistability occurs below the line $\delta =\delta_b$.
The curve $P(E_H^-)=P(E_b^-)$ (black solid) divides the bistable region into a domain labeled BI+SP where SP
sets in only for powers above the bistable knee for up-switching, and a domain labeled SP, 
where stable up-switching is not possible, being hampered by SP, which dominates the dynamics.} 
\label{fig:map2D}\end{figure}

The analysis reported above shows that the bistable response depends only on the detuning, while the time constant $\tau$ can affect 
qualitatively the onset of SP due to the Hopf bifurcation. In fact different scenarios are possible depending on the existence
of one or both Hopf thresholds (in turn corresponding to the roots in Eq.~\eqref{eq:Hopf}) being real),
and whether such threshold occurs at powers above or below the bistable knee for up-switching.
Four possible scenarios are displayed in Fig.~\ref{fig:eigs}, 
where we show the bistable stationary response along with the unstable eigenvalues responsible for instabilities. 
First, Fig.~\ref{fig:eigs} shows the well-known fact that a purely real and positive eigenvalue of $M$ leads to instability of the steady state solution along the negative slope branch, 
which turns out to be a saddle point in phase-space.
Conversely, SP is characterized by a pair of complex conjugate eigenvalues,
and the relative threshold are highlighted by empty triangles. We find such threshold to occur always on the upper branch
of the bistable response (or when the response is monotone, see below). However, depending on the value of $\tau$, 
the lower Hopf threshold $P(E_H^-)$ can take place above [as in Fig.~\ref{fig:eigs}(a,c)] or below [see Fig.~\ref{fig:eigs}(b,d)] the knee $P(E_b^-)$
which characterize the bistable up-switching. In the former case the cavity can exhibit bistable up-switching to a stable steady state,
whereas in the latter case up-switching occurs inevitably towards a SP-unstable state: stabilization requires to move along the hysteresis cycle below the Hopf threshold.
Moreover, also depending on the value of $\tau$, the upper branch can be indefinitely SP-unstable above the first threshold $P(E_H^-)$ [see Fig.~\ref{fig:eigs}(a,b)], 
or, viceversa, exhibit a SP switch-off energy or secondary threshold beyond which SP ceases to take place [see Fig.~\ref{fig:eigs}(c,d); 
note that we change the detuning in Fig.~\ref{fig:eigs}(d) only to make the picture clearer, though the same qualitative behavior occurs at $\delta=-4$].
Indeed, if $\tau \geq 1$, Eq.~\eqref{eq:Hopf} may also admit a real solution $E_H^{+}$, and hence SP occurs only in a finite
range of energies (and powers) $E_H^-<E<E_H^+$. When $\tau$ grows, this finite interval shrinks, 
and vanishes as a critical value $\tau=\tau_c$ is reached.
For $\tau=\tau_c$, SP no longer takes place and the entire upper branch becomes stable.
From Eq.~\eqref{eq:Hopf}, we find such critical value $\tau_c$ to be given by the positive real root of the cubic polynomial
(its explicit expression is too cumbersome) 
\begin{equation}\label{eq:tauc}
\tau_c^3 + \tau_c^2 - \left(1+\frac{\delta^2}{4} \right) \tau_c -1 = 0,
\,\,\tau_c\approx\frac{|\delta|}{2},\,\,|\delta|\gg1.
\end{equation}

The behavior discussed above can be clearly seen by reporting the SP threshold energies $E_H^{\pm}$
versus $\tau$, at constant detuning. Such plot, displayed in Fig.~\ref{fig:th} for $\delta =-4$, 
shows a shaded region ($E_H^-<E<E_H^+$), which correspond to SP. 
We clearly see that no SP occurs for $\tau > \tau_c$.
Conversely, decreasing $\tau$ below $\tau_c$ results into widening the portion of the upper branch that exhibits SP,
until below $\tau=1$ the whole upper branch above the switch-on threshold $E_H^-$ turns out to be unstable. 
In this case, the SP switch-off energy $E_H^+$ diverges as the asymptote $\tau=1$ (dashed vertical line) is approached.  
Importantly for $\tau \rightarrow 0$ also the first threshold $E_H^{-}$ diverges, 
which means that a finite response time is a key ingredient for SP to be observable. 
This is consistent with the fact that Kerr instantaneous nonlinearities yield no SP at all.
In fact, in this case, the eigenvalues are easily found to be $\lambda^{\pm}=-1 \pm \sqrt{E^2 - (\delta + 2\chi E)^2}$, 
which rule out the possibility to have a complex conjugate pair with positive real part.

\begin{figure}[h!] 
\centering
\includegraphics[width=8.0cm]{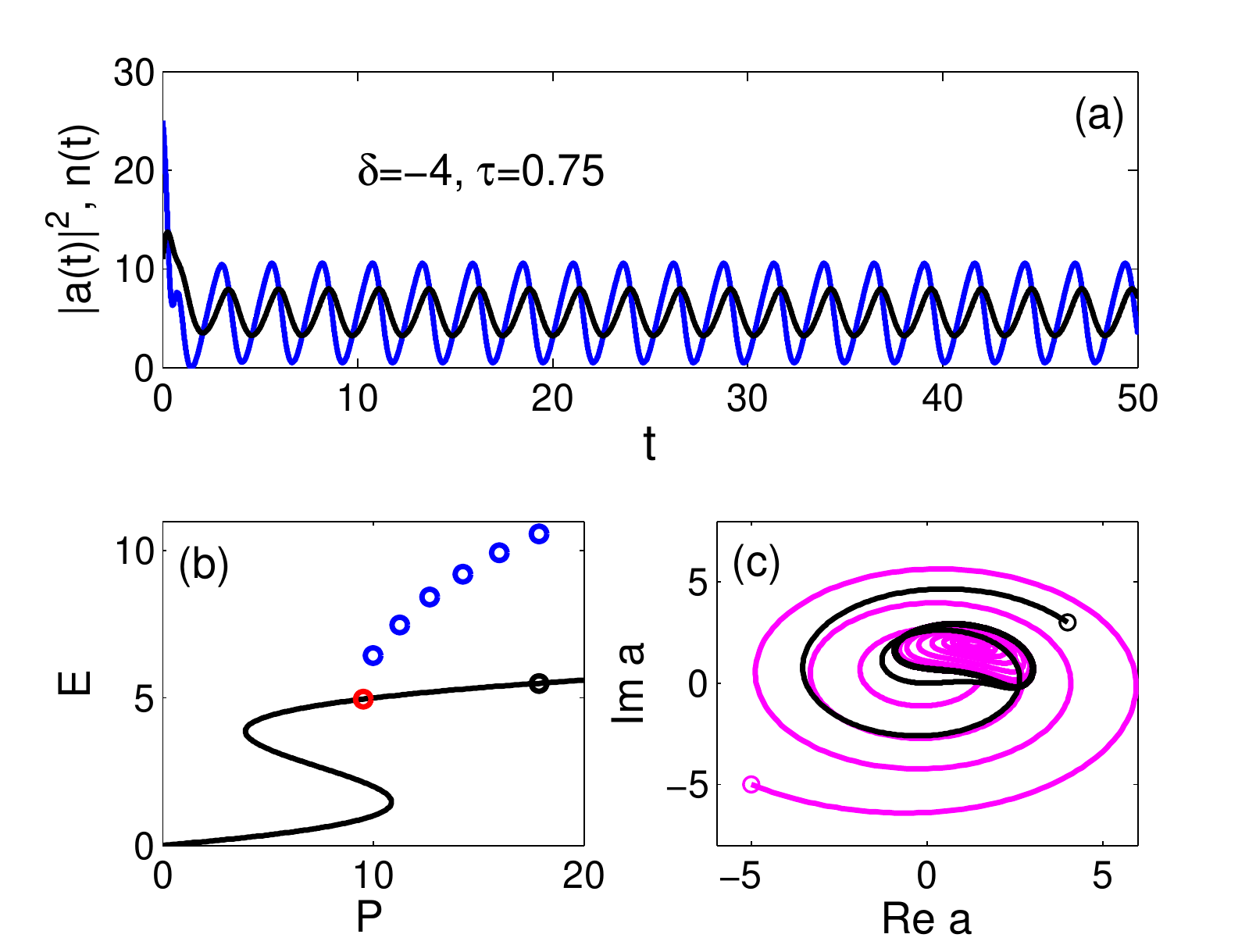}
\caption{(Color online) Dynamics of SP ruled by Eqs. (\ref{eq1}-\ref{eq2}), with $\delta = -4$, and $\tau =0.75$:
(a) temporal evolution of the intra-cavity energy and carrier density corresponding to the rightmost blue circle in (b); 
(b) steady response with superimposed peak energy of the periodic oscillations (blue open circles); The red filled circle marks
the Hopf bifurcation point. The black filled circle marks the SP-unstable steady state which gives rise to the dynamics shown in (a,c);
(c) phase-space picture of the optical field showing the attracting limit cycle from two different initial conditions (open circles);
} 
\label{dynamics}\end{figure}

In order to have a complete picture and further show how the onset of SP changes with detuning, 
we have drawn in Fig.~\ref{fig:map2D} a color map of the level curves of SP threshold energies $E_H^{\pm}$, 
in the parameter plane ($\tau, \delta$). A number of interesting observations can be drawn.
The SP region is bounded by the border $\tau=\tau_c$, and in the bistable region ($\delta<\delta_b$)
$\tau_c$ decreases with decreasing values of absolute detuning $|\delta|$.
Interestingly enough, the scenario illustrated in Fig.~\ref{fig:th} remains valid also for detunings $\delta > \delta_b$,
where bistability disappears. Finally in the region of positive detunings, we are left with the upper branch being
fully unstable for all energies $E>E_H^-$. In this region, the threshold energy $E_H^-$ diverges, not only in the instantaneous limit $\tau=0$,
but also for $\tau=1$, whereas relatively low values of $E_H^-$ are found for $\tau \sim 1/2$, 
i.e. when the response time of the nonlinearity is nearly equal to the photon lifetime.
Furthermore, the bistable region is divided into two distinct domains by the (solid black) curve 
which arises from the condition $P(E_H^-)=P(E_b^-)$ (its explicit expression is too cumbersome).
In the domain BI+SP above such curve (bounded from above also by the line $\delta=\delta_b$), 
one has that the cavity can work as a bistable switch since
the upper branch right above the knee for up-switching is stable  [as in Fig.~\ref{fig:eigs}(a,c)], 
whereas in the domain labeled SP below the curve,
up-switching is no longer allowed, since the upper branch above the knee is SP-unstable [as shown in Fig.~\ref{fig:eigs}(b,d)]
The reader can easily recognize a qualitative similarity of the picture discussed here 
with the dynamics of SP ruled by free carrier dispersion, recently discussed in Ref. \cite{Malaguti11}.
Although a detailed analytic investigation of the stability of the SP-oscillating state (limit cycle) is beyond the scope of this paper,
similarly to the case discussed in Ref.~\cite{Malaguti11}, our numerical simulations of Eqs.~\eqref{eq:model} 
suggest that stable limit cycles, working as attractors from a large basin, 
exist in a wide domain of the parameter plane (witnessing the supercritical nature of the Hopf bifurcation). 
An example of such stable dynamics is shown in Fig.~\ref{dynamics}. 

{Let us comment on the observability of the SP dynamics. In nanocavities with high $Q$ ($Q=10^3-10^5$) the photon lifetime
in the near infrared ranges from few picoseconds to tens of picoseconds. While the constraint to have a response time of the nonlinearity
of the same order of magnitude is naturally met in semiconductors with nonlinear response dominated by free-carrier dispersion,
the same constraint in the framework of the Kerr model rule out the possibility to observe SP dynamics
in media with nonlinearities of electronic origin since they are too fast (fs range). 
Nevertheless the predictions of our Kerr model become interesting for Kerr-like materials with response time
in the ps range such as, e.g., soft matter, metal films \cite{Conforti11}, or more traditional liquids with reorientational nonlinearity,
which are still the object of recent studies \cite{Fanjoux08,Conti10}. In particular, for instance, highly nonlinear liquids such as nitrobenzene \cite{nitrobenzene} or CS$_2$ could be easily employed to fill a photonic crystal matrix (as also recently proposed for microstructured fibers \cite{Conti10}), while metal films could be employed in conjunction with dielectrics to form a single cavity or cavity arrays \cite{Sipe99}.
Assuming, for instance, a response time $\tau_{phys} \sim 30$ ps \cite{nitrobenzene}, 
which yield $\tau =(\tau_{phys}/2t_c)\sim 0.75$  in a cavity with $Q \sim 25000$
($t_c \sim 20$ ps at $\lambda=1.55 \mu$m), assuming $n_{2I} \sim 10^{-17} m^2/W$ and a nonlinear modal volume $V=3 (\lambda/n)^3$, 
the threshold power $P=10$ in Fig.~\ref{dynamics} corresponds to a real-world power 
$P_{in}=(\gamma/\Gamma_0^2) P \sim 10$ mW in the waveguide coupled to the nanocavity, 
where $\gamma=\omega_0 n_{2I} c/(n_{eff}nV)$ is the overall nonlinear coefficient  \cite{deRossi09}.
Here we have assumed a refractive index $n \sim n_{eff} \sim 1.5$ and $Q$ to be essentially determined by the coupling itself.
}

Having characterized so far the threshold for SP and its competition with bistability, 
since Ikeda and Akimoto have shown that the limit cycles destabilize, leading eventually to chaos \cite{Ikeda82},
in the next section we deepen this point with the aim of determining 
the domain of the parameter plane where the transition to chaos could be observed.
\section{The turbulent regime}

In Ref.~\cite{Ikeda82} Ikeda and Akimoto have studied the transition to chaos, identifying a period doubling cascade up to $2^2P$ (i.e.~oscillation with period-four) 
at a fixed value of detuning. Here we report further details about the emergence of chaos in a wide domain of  parameters.
We employ different tools, ranging from  Poicar{\'e} section and its corresponding bifurcation diagram to the calculation of Lyapunov exponents.
Our principal purpose is to assess whether a nano-cavity described by the model~\eqref{eq:model} can work as a reliable bistable switch,
and hence whether the onset of chaos should be expected when the cavity operates progressively off-resonance,
especially in the region labeled BI+SP in Fig.~\ref{fig:map2D}.
\begin{figure}
	\centering
		\includegraphics[width=0.235\textwidth]{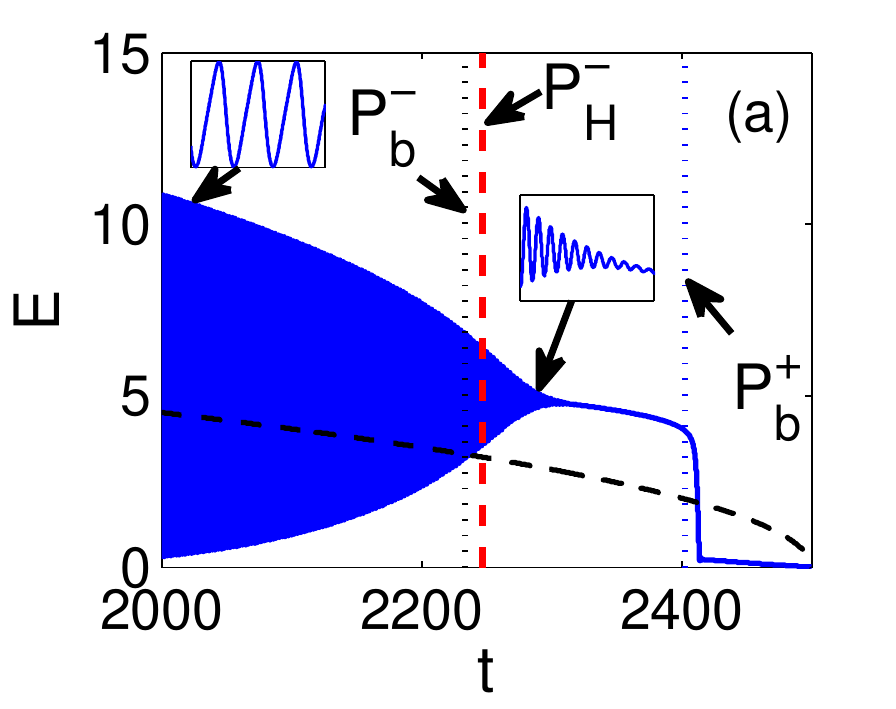}
		\includegraphics[width=0.235\textwidth]{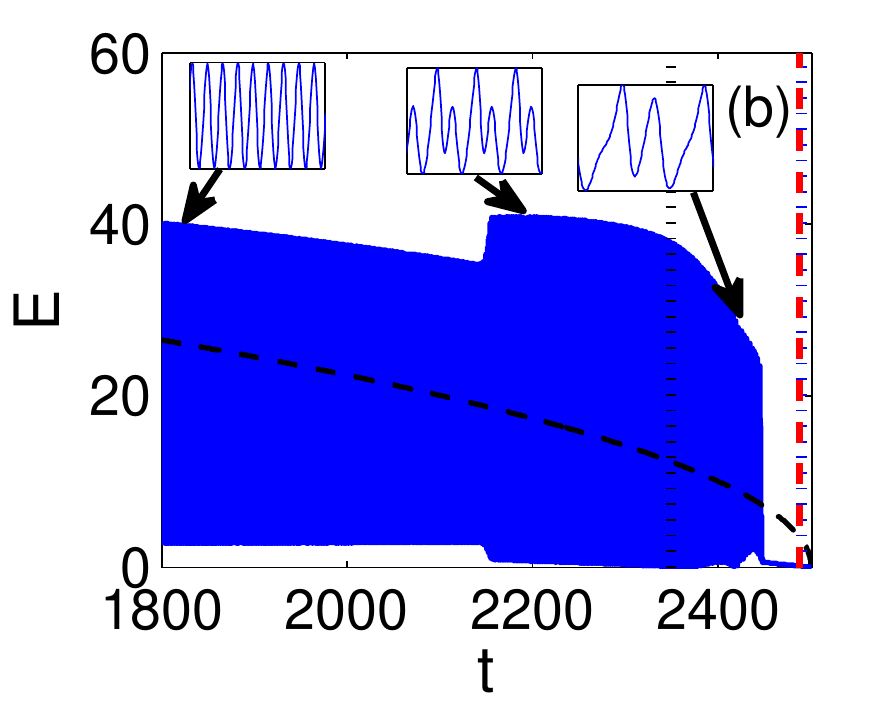}		
	\caption{(Color online) Dynamical evolution ruled by Eqs.~\eqref{eq:model} as 
	the forcing term $\sqrt{P}$ varies adiabatically in time (dashed line). Here $\tau = 0.45$, and (a) $\delta=-4$; (b) $\delta=-10$. 
	The vertical lines mark the time instants at which the input power cross the main bifurcation points: 
	bistable knees $P_b^\pm$ (dotted blue and black, respectively) and Hopf threshold $P_H^-$ (dashed red). 
	The insets  zoom over characteristic time intervals, indicated by arrows. 
	Notice that in (b) the two thresholds $P_b^\pm$ and $P_H^-$ almost overlap.}
	\label{fig:Ebifurcation_tau045_deltam4}
\end{figure}
\begin{figure}
	\centering
		\includegraphics[width=0.45\textwidth]{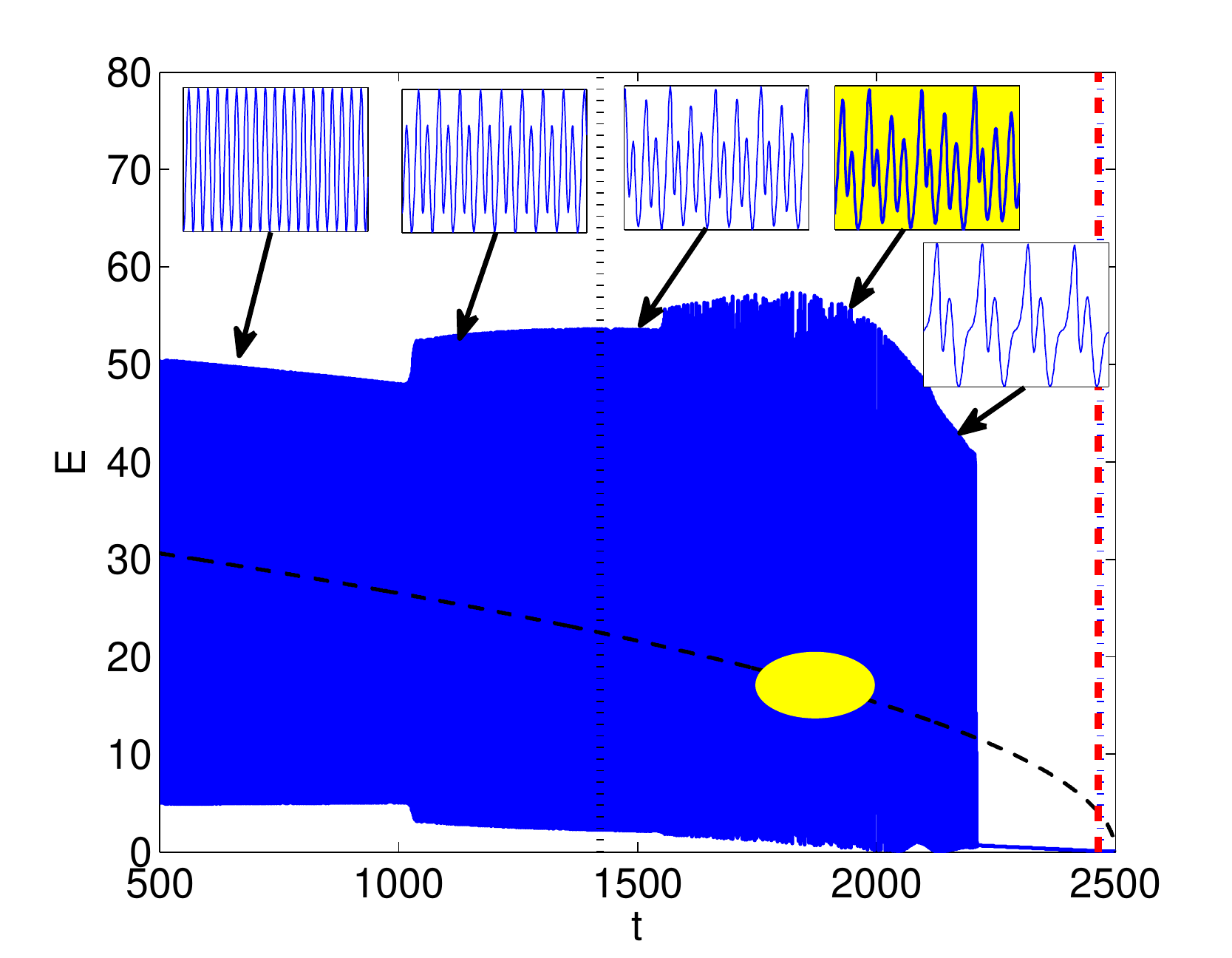}
	\caption{(Color online) As in Fig.~\ref{fig:Ebifurcation_tau045_deltam4}, with $\delta=-15$.
	The chaotic region is highlighted in yellow.}
	\label{fig:Ebifurcation_tau045_deltam15}
\end{figure}
To begin with, it is instructive to report about the dynamics ruled by Eqs.~\eqref{eq:model} when, starting above the threshold $P_H^-$,
the input power $P$ is adiabatically decreased. In fact, this is the situation where the onset of chaos is expected according to Ref. \cite{Ikeda82}.
We start at moderately low detuning ($\delta=-4$) and for $\tau=0.45$, which corresponds to SP being unbounded on the upper branch.
As shown in Fig.~\ref{fig:Ebifurcation_tau045_deltam4}(a), in this case, the Hopf bifurcation is clearly supercritical, 
since the system settles on a limit cycle, whose amplitude vanishes approaching the bifurcation point  (approximately as $[E-E_{H}^-]^{1/2}$).
However, at larger (in modulus) detunings, we observe a jump in $|a|^2$ as the limit cycle ($1P$) looses its stability and the system settles on a period-two ($2P$) oscillation,
as shown in Fig.~\ref{fig:Ebifurcation_tau045_deltam4}(b) for $\delta=-10$. The $2P$ solution does not visit anymore the simplest 1P limit cycle. 
Conversely it abruptly switches to the stable low-branch steady state. Remarkably this happens still above the Hopf bifurcation point $E_H^-$.
A more complex sequence of period doubling bifurcations and chaotic motion are detected at higher detunings when the input power approaches the knee value $P_b^-$. 
In Fig.~\ref{fig:Ebifurcation_tau045_deltam15}, where $\delta=-15$, oscillations with several different periods are evident. 
Moreover a chaotic regime appears, in the range $\sqrt{P} \approx17-20$ ($P\approx280-400$). 
In phase space this corresponds to the appearance of a strange attractor (not shown because its structure is already illustrated in Ref. \cite{Ikeda82}).

\begin{figure}
	\centering
		\includegraphics[width=0.45\textwidth]{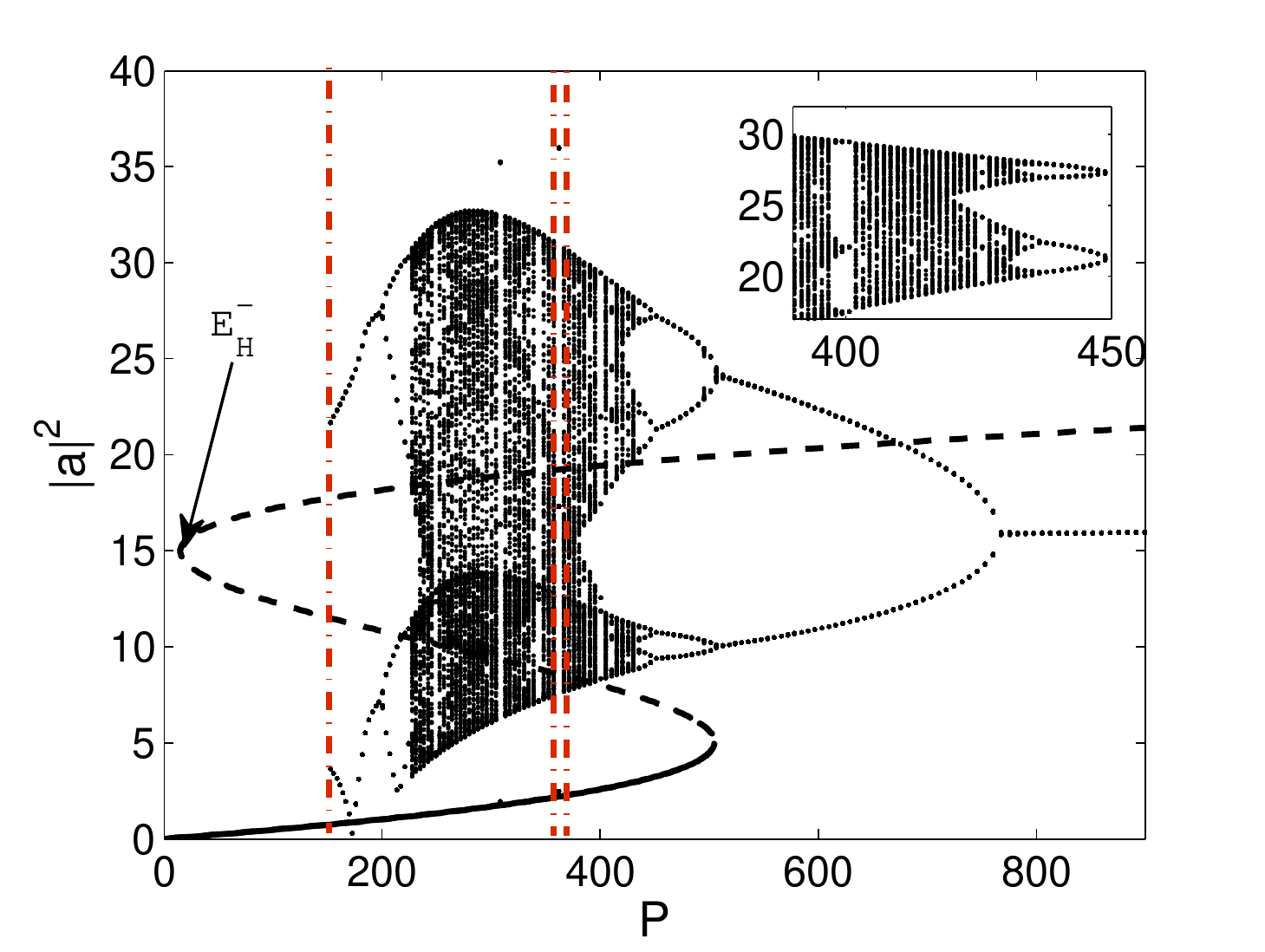}
	\caption{(Color online) Bifurcation diagram, $\tau=0.45,\,\,\delta=-15$. The vertical dashed lines highlight a $3P$ window and the collapse at small $P$. Inset shows a detail in which period doubling and periodic windows can be identified more clearly.}
	\label{fig:bifdiagram}
\end{figure}

Since following simply the adiabatic dynamics could be possibly misleading (e.g.~because of critical slowing down), 
as it neglects the rich variety of phenomena that occurs over the small scale, 
we have drawn also bifurcation diagrams calculated by collecting trajectory points on a Poincar{\'e} section for different powers.
A typical example, using the same parameters as in Fig.~\ref{fig:Ebifurcation_tau045_deltam15},
and defining the Poincar{\'e} section on the fixed phase $\angle{a}-\pi=0$ of the intra-cavity field, 
is reported in Fig.~\ref{fig:bifdiagram}. We can clearly identify a period doubling cascade (up to $2^3P$), as well as chaotic regimes.
The onset of chaos follows a non-trivial scenario where narrow windows of period-three solutions ($3P$) are interspersed between two ranges
of powers where the motion turns out to be chaotic. The vertical dashed lines in Fig.~\ref{fig:bifdiagram} mark indeed a $3P$ window. 
This is analogous to the Feigenbaum's route to chaos and confirms the observation of chaos for $P\approx220-380$, already drawn above from Fig.~\ref{fig:Ebifurcation_tau045_deltam15}.

The bifurcation diagram is computed up to $P\approx 150$ because lower power levels do not result into any limit cycle.
Vice-versa the solution is observed to collapse toward the stable node represented by the lower branch of the bistable response.
This phenomenon is independent from chaos, as mentioned above with reference to detuning $\delta = -10$. 
 It occurs at large negative detunings for input powers above the SP threshold ($P>P_H^-$). 
This can be qualitatively explained by the coexistence of a stable fixed point (lower branch solution), a saddle (negative-slope branch), and an unstable limit cycle.
As the $2P$ limit cycle spans the phase space with wide oscillations around the upper branch, it can approach the saddle point 
(which near the first bistable knee is closer in phase space to the center of the oscillations) being forced away,
until eventually it can be captured by the lowest energy (stable) solution.

\begin{figure}
	\centering
		\includegraphics[width=0.45\textwidth]{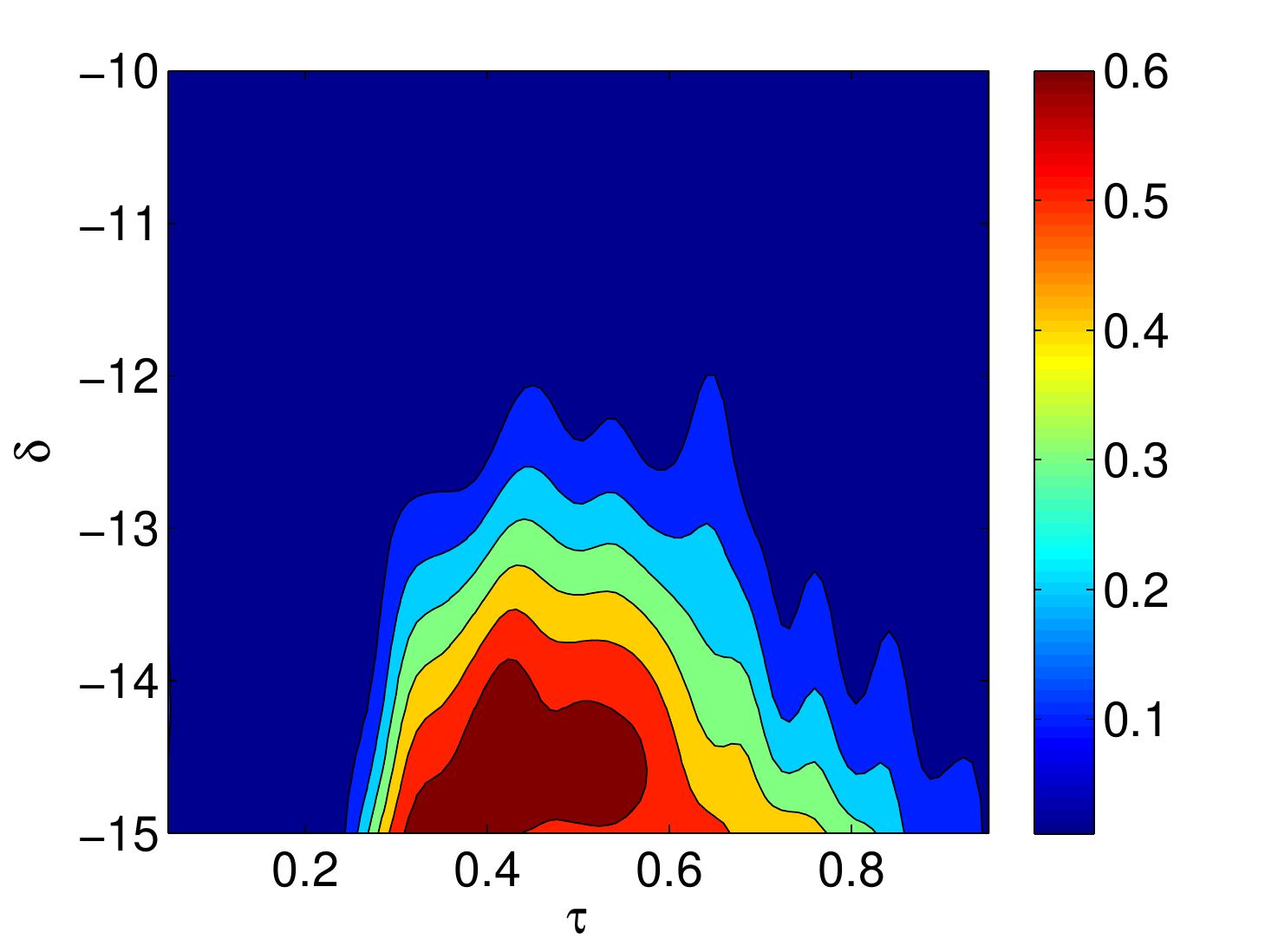}
	\caption{(Color online) Color level map of maximal Lyapunov exponent in the parameter plane $(\tau,\delta)$. }
	\label{fig:Lyapunov}
\end{figure}

While the bifurcation map is a useful visual tool to characterize the onset to chaos and the full dynamics at fixed parameters,
in order to explore in which region of the parameters one should expect to observe the chaotic dynamics, we have resorted 
to compute the dominant (maximal) Lyapunov exponent.  We have explored a wide region of the parameter plane $(\tau,\delta)$, 
where, in each point of such plane, we have iterated over the values of power $P$ to find the largest exponent. 
We recall that a positive Lyapunov exponent  (within numerical inaccuracies) entails that the system exhibits a chaotic behavior
(here quasi-periodic motion is excluded by the dissipative character of our model ~\eqref{eq1}).
From the map displayed in Fig.~\ref{fig:Lyapunov}, we notice that chaotic motion manifests itself only
when the cavity is detuned far off-resonance (i.e. at very large values of $|\delta|$), 
provided that the SP-unstable range is not finite, or in other words that the Hopf bifurcation is not bounded from above ($E_H^+\to \infty$)
which requires $\tau<1$. 
Therefore we can draw the important conclusion that the region where the cavity could work as a bistable switch
is mutually exclusive with chaos. Therefore the onset of chaos cannot spoil the behavior of the cavity as a switch,
once the latter is used in the region labeled BI+SP in Fig.~\ref{fig:map2D}.
{ We point out that, in terms of power, the observation of chaos is much more challenging than stable SP since  power levels leading to the former turn out to be much larger than those leading to the latter; indeed at very large detuning, which corresponds to several times the cavity linewidth, bistability is observed at much higher power level (compare the horizontal axis scale in Fig.~\ref{dynamics} and Fig.~\ref{fig:bifdiagram}).  
}

As a final remark about the existence of $2^nP$ periodic solutions, we point out that they can be detected only when $P_H^-<P_b^-$ 
[as in the examples shown in Fig.~\ref{fig:eigs}(b,d)]. 
As discussed with reference to Figs.~\ref{fig:eigs}-\ref{fig:map2D}, this may occur not only for both $\tau<1$ (as shown explicitly above), 
but also for $\tau>1$, where the Hopf bifurcation is bounded from above.  
More specifically $2P$ solutions are easily seen in a small subset of the region marked as SP in Fig.~\ref{fig:map2D}.
In this case two unstable solutions, namely a repulsive (negative slope) branch and a SP branch, coexist. 
This seems to be a key ingredient for the limit cycles to loose their stability. 

\section{Conclusions}
In this work we have revisited the model that rules the behavior a passive {\em small} cavity with Kerr delayed response, 
pioneered in Ref.~\cite{Ikeda82}. We have reported a full characterization of SP instabilities and their competition with bistability, 
outlining the existence of different possible scenarios. Importantly we have found a maximal critical value for the relaxation time that allows SP to occur,
and have shown that SP can have two bifurcation points, while it can occur also in the absence of bistability.
We have further characterized the destabilization mechanism of the limit cycle in the full parameter space, finding  that chaos is mutually exclusive with
the domain where the cavity can be employed as a bistable switching element.  
{In particular the chaotic regime predicted by Ikeda \cite{Ikeda82}
in this system requires to be detuned strongly  off-resonance,  in turn implying the use of extremely high powers, 
thus making its observation rather challenging.
Viceversa, in contrast with Kerr instantaneous nonlinearities, the observation of stable SP appear feasible in high-Q nanocavities 
filled with Kerr-like media with response time in the range of picoseconds.
}
Future work will be devoted to study the effect of coupled cavity systems and the interplay of different nonlinear mechanisms.
\acknowledgments
This work was supported by the European Commission, in the framework of the Copernicus project (no.~249012).

\end{document}